\documentclass[pdftex,twocolumn,epjc3]{svjour3}        

\RequirePackage[T1]{fontenc}

\smartqed  

\RequirePackage{graphicx}
\RequirePackage{mathptmx}      
\RequirePackage{flushend}
\RequirePackage[numbers,sort&compress]{natbib}
\RequirePackage[colorlinks,citecolor=blue,urlcolor=blue,linkcolor=blue]{hyperref}
\usepackage{siunitx}
\usepackage{textcomp}

\journalname{Eur. Phys. J. C}

\begin{document}

\title{muCool: A next step towards efficient muon beam compression}

\author{
		A. Antognini\thanksref{addr1,addr2}
	\and
		Y. Bao\thanksref{addr2}
	\and
		I. Belosevic\thanksref{e1,addr1}
	\and
		A. Eggenberger\thanksref{addr1}
	\and
		M. Hildebrandt\thanksref{addr2}
	\and
		R. Iwai\thanksref{addr1}
	\and
		D. M. Kaplan\thanksref{addr3}
	\and 
		K. S. Khaw\thanksref{addr1,addr5}
	\and
		K. Kirch\thanksref{addr1,addr2}
	\and
		A. Knecht\thanksref{addr2}
	\and
		A. Papa\thanksref{addr2,addr6}
	\and
		C. Petitjean\thanksref{addr2}
	\and
		T. J. Phillips\thanksref{addr3}
	\and
		F. M. Piegsa\thanksref{addr1, addr4}
	\and
		N. Ritjoho\thanksref{addr2}
	\and
		A. Stoykov\thanksref{addr2}
	\and
		D. Taqqu\thanksref{addr1}
	\and
		G. Wichmann\thanksref{addr1}
	}

\thankstext{e1}{e-mail: ivanabe@phys.ethz.ch}

\institute{Institute for Particle Physics and Astrophysics, ETH Z\"urich, 8093 Z\"urich, Switzerland\label{addr1}
          \and
          Paul Scherrer Institute, 5232 Villigen-PSI, Switzerland\label{addr2}
          \and
		  Illinois Institute of Technology, Chicago, IL 60616 USA\label{addr3}
		  \and
		  Dipartimento di Fisica, Universit\`a di Pisa, and INFN sez. Pisa, Largo B. Pontecorvo 3, 56127 Pisa, Italy\label{addr6}
		  \and
		  \textit{Present address}: Department of Physics,University of Washington, Seattle, WA 98195, USA\label{addr5}
		  \and
		  \textit{Present address}: Laboratory for High Energy Physics, Albert Einstein Center for Fundamental Physics, University of Bern, CH-3012 Bern, Switzerland\label{addr4}
}

\date{Received: date / Accepted: date}

\maketitle

\begin{abstract}
	A novel device to compress the phase space of a muon beam by a factor of $10^{10}$ with a $10^{-3}$ efficiency is under development. A surface muon beam is stopped in a helium gas target consisting of several compression stages, wherein strong electric and magnetic fields are applied.  The spatial extent of the stopped muon swarm is decreased by means of these fields until muons with eV energy are extracted into vacuum through a small orifice. It was measured that a 20~\si{\centi \meter} long muon stop distribution can be compressed in longitudinal direction to sub-mm extent within 2~\si{\micro \second}.
	 Additionally, a drift perpendicular to the magnetic field of the compressed low-energy muon swarm was successfully demonstrated, paving the way towards the extraction from the gas and re-acceleration of the muons. 
\end{abstract}

\section{Introduction \label{SecIntroduction}}

Standard surface muon beams have a relatively poor phase space quality. An improvement of the muon beam quality would open the way for new experiments in low energy particle physics, atomic physics and material research. Examples of such experiments include muonium 1S-2S spectroscopy, muon electric dipole moment (EDM) measurement and muon spin rotation ($\mu$SR) applications. Due to the limited lifetime of the muons ($\tau=2198$~\si{\nano \second}), a fast cooling scheme is required, and thus conventional beam cooling methods such as stochastic~\cite{VanderMeer1985} or electron cooling~\cite{Budker1978} cannot be applied.

A new phase space compression scheme (muCool device) has been proposed \cite{Taqqu2006}, which transforms a standard surface muon beam into a beam of high brightness and low energy. Within about 10~\si{\micro \second}, the incoming $\mu^+$ are stopped in He gas, manipulated by electric and magnetic fields in order to decrease their spatial extent, and extracted into vacuum again. The 6D phase space is reduced by a factor of $10^{10}$ with an efficiency of $10^{-3}$, mainly limited by the muon lifetime. Thus, the brightness of the incoming beam is enhanced by a factor of $10^{7}$.  

The decrease of the muon swarm extension is obtained by making the drift velocity of the $\mu^+$ in the gas position-dependent. The drift velocity vector of $\mu^+$ in gas, in the presence of electric and magnetic fields, can be written as \cite{Blum}:
\begin{equation}
\vec{v}_D = \frac{\mu |\vec{E}|}{1 + \omega^2 / \nu^2} \left[ \vec{\hat{E}} + \frac{\omega}{\nu} \vec{\hat{E}} \times \vec{\hat{B}} + \frac{\omega^2}{\nu^2} \left(\vec{\hat{E}} \cdot \vec{\hat{B}} \right) \vec{\hat{B}} \right], 
\label{EqDriftVelocity}
\end{equation}
where $\mu=e/\nu m$ is the mobility of the muon, $\vec{\hat{E}}$ and $\vec{\hat{B}}$ are the unit vectors of the electric  and  magnetic fields, $\omega = eB/m$ is the cyclotron frequency of the muon and $\nu$ the average collision frequency of the muon with the He gas atoms. 
The drift velocity $\vec{v}_D$ can be made position-dependent by applying a position-dependent electric field $\vec{E}$ and/or by making the collision frequency $\nu$ position-dependent, which changes the weights of the three components in Eq.~\ref{EqDriftVelocity}.

\begin{figure}
	\includegraphics[trim=0 0 0 0cm,clip=true,width=1.0\columnwidth]{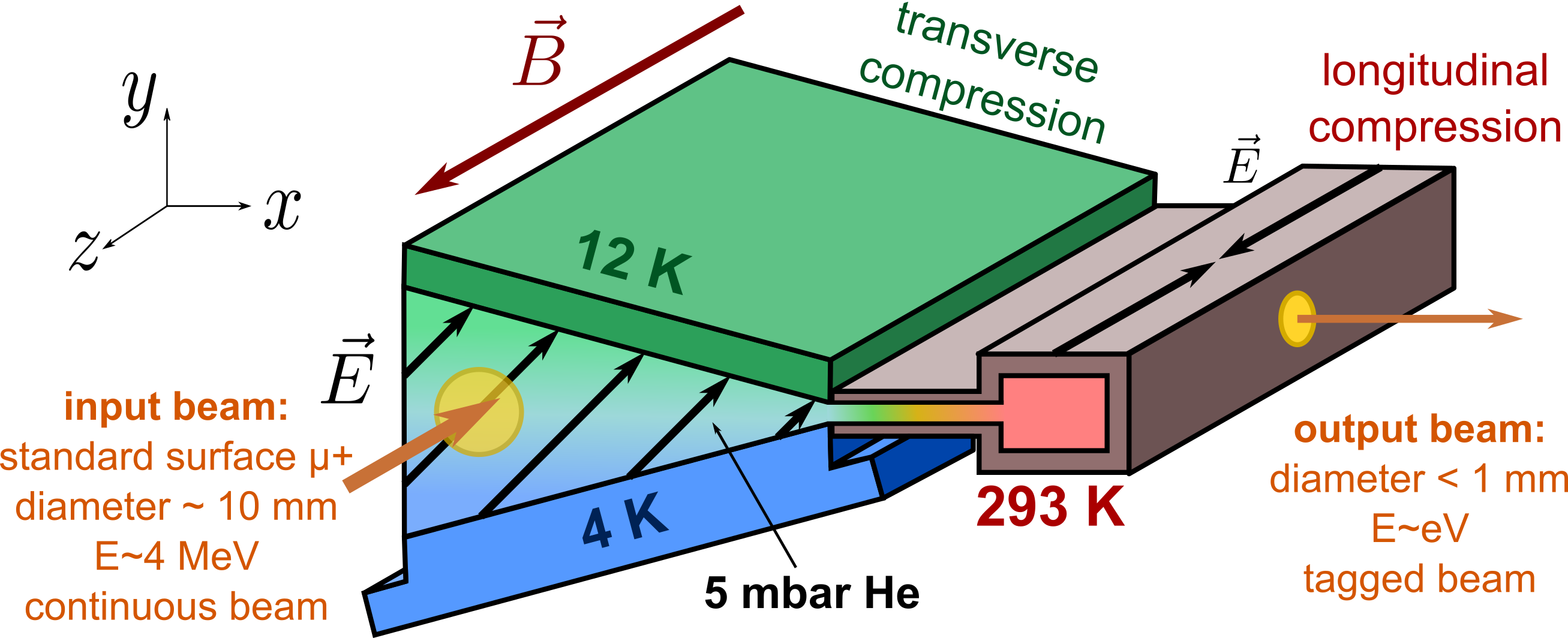}%
	\caption{\label{FigWorkingPrinciple}Overview of the proposed phase space compression scheme. A standard surface muon beam is stopped in a cryogenic helium gas with a vertical temperature gradient inside a 5~T magnetic field. The stopped muon swarm is then compressed by means of $\vec{E}$- and $\vec{B}$-fields in the $y$-direction (``transverse compression''), and at the same time it is steered in $x$-direction towards the longitudinal compression stage. There, the He gas is at room temperature, and there is a component of the electric field parallel to the magnetic field axis. Thus, the muon swarm is compressed towards the center of this stage. Finally, the muons with a swarm size of $\mathcal{O}$(mm) are extracted through a small orifice into vacuum, where they are re-accelerated and sent to another experiment.}  
\end{figure}

A sketch of the He gas target where such a compression occurs is shown in Fig.~\ref{FigWorkingPrinciple}, together with the coordinate system used throughout this paper. The target consists of several stages and is placed in a homogeneous magnetic field, oriented along the $+z$-direction, i.e. $\vec{B}=(0,0,B)$, with $B=5$~\si{Tesla}.

The incoming $\mu^+$ beam is stopped in the first stage that is at cryogenic temperatures, with a vertical ($y$-direction) gas density gradient induced by a temperature gradient between $T=4$~\si{\kelvin} at the bottom and $T=12$~\si{\kelvin} at the top of the roughly 3~\si{\centi \meter} high target \cite{Wichmann}. The gas density gradient gives rise to a position-dependent collision frequency $\nu=\nu(x,y,z)$.

An electric field $\vec{E}=(E_x,E_y,0)$ with $E_x=E_y \approx 1~\si{\kilo \volt \per \centi \meter}$  is applied perpendicular to the magnetic field in this stage, thus only the first two terms in Eq.~\ref{EqDriftVelocity} remain. The muons in a region of higher gas density (colder temperatures) follow the $\vec{\hat{E}}$-direction because $\nu$ is large, whereas muons in a region of lower gas density (warmer temperatures) primarily follow the $\vec{\hat{E}}\times\vec{\hat{B}}$-direction. With our choice of $\vec{E}$, $\vec{B}$ and $\nu(x,y,z)$ we achieve compression of the stopped muon swarm in vertical ($y$) direction (called ``transverse compression"), with superimposed drift in $+x$-direction towards the second compression stage, where longitudinal (in $z$-direction) compression occurs.

The second stage is at room temperature (much lower gas density) and the electric field is parallel to the magnetic field: $\vec{E}=(0,0,\mp E_z)$ pointing towards the center of the target, as indicated in Fig.~\ref{FigWorkingPrinciple}, with $E_z\approx50$~\si{\volt \per \centi \meter}. Therefore, in the second stage Eq.~\ref{EqDriftVelocity} simplifies to $\vec{v}_D \approx \mu \vec{E}$, and the muon swarm is compressed along the $\pm z$-direction (``longitudinal compression") towards $z=0$. Adding an electric field in the $y$-direction gives rise to a superimposed drift in $+x$-direction, which guides the compressed muon swarm towards an orifice of about 1~mm diameter, from where the muons are extracted into vacuum. After the extraction into the vacuum, the $\mu^+$ are accelerated to keV energy, while keeping eV energy spread and are successively extracted from the magnetic field. 

Initially, the various stages of the compression target can be tested separately, simplifying the experimental approach considerably. Transverse compression has been successfully demonstrated and will be presented in a separate publication. Longitudinal compression has been demonstrated to be feasible~\cite{Bao2014}. 
However, several problems prevented quantification of the muon swarm compression efficiency. The major issue was impurities in the helium gas target, which captured low-energy muons and thus lowered the compression efficiency. Moreover, a large background allowed to demonstrate only qualitatively the feasibility of the longitudinal compression.

In this paper we present an improved demonstration of the longitudinal compression with a  quantification of the compression efficiency. 
Furthermore, we demonstrate that the muon swarm can be steered in $+x$-direction by adding a vertical ($y$-direction) electric field. This is crucial to guide the $\mu^+$ through the different compression stages and finally towards the point of extraction. The experiments were performed at the $\pi$E1 beam line at the Paul Scherrer Institute (PSI), which delivers about $10^4$ $\mu^+$/s at $11$~MeV/c.

\begin{figure} []
	\includegraphics[width=\columnwidth]{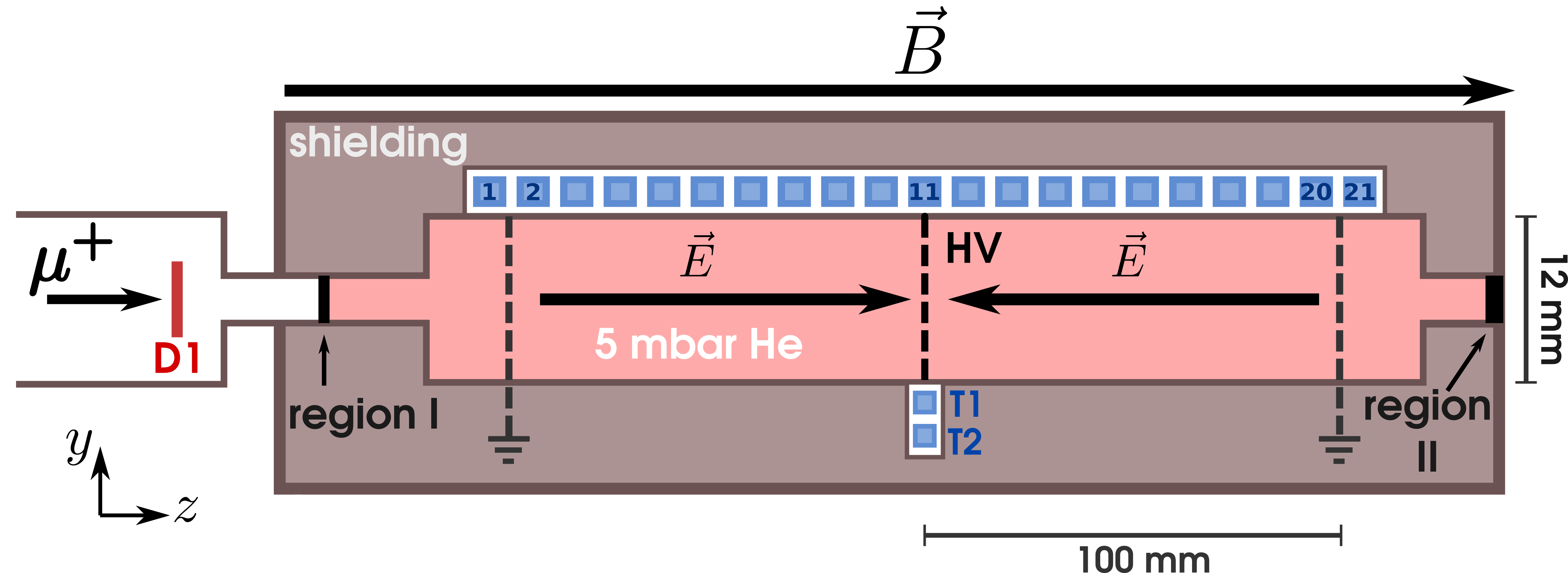}%
	\caption{\label{FigSetup2014}Sketch of the He gas target used to measure the longitudinal compression (not to scale). The muon beam passes through an entrance detector D1, target window (region I) and enters the He gas. Some of the muons are stopped in the gas and subjected to the electric field, pointing towards the center of the target. A large fraction of the muons passes through the target and stops in region II. Several scintillators (S1, $\ldots$, S21, T1, T2) detect the positrons from muons decaying in their vicinity.}
\end{figure}

\section{Test of longitudinal compression \label{SecSetup}}
In order to test longitudinal compression, an 11~MeV/c muon beam was injected longitudinally into a target containing He gas of a few mbar pressure at room temperature. The setup is sketched in Fig.~\ref{FigSetup2014}. The muons first have to pass through a 55 \si{\micro \meter} thick entrance detector (D1), giving the initial time $t_0 = 0$, then through a 2~\si{\micro \meter} Mylar target window enclosing the He gas. Only a small fraction $\mathcal{O}(1\%)$ of the muons producing the signal in D1 stop in the gas, while the remaining muons stop in D1 or target window (region I in Fig.~\ref{FigSetup2014}) or downstream part of the target (region II in Fig.~\ref{FigSetup2014}).

The longitudinal compression target had a transverse cross section of $12 \times 12$~\si{\square \milli \meter}, and an ``active" length, where the electric field was defined, of about 200~\si{\milli \meter}. Side walls of the target were lined with gold electrodes that created a V-shaped electric potential with a minimum at the center of the target cell at $z=0$. This choice of the electric potential produced an electric field pointing towards $z=0$, which caused the muons to move along the $\pm z$-direction towards the potential minimum \footnote{Because of $\nabla \cdot E=0$, at $z=0$ the electric field will also have a radial component. This component is included automatically in the COMSOL Multiphysics\textregistered~simulation but it is practically so small that it does not affect significantly the muon motion.  }.

The exact electric field was simulated using a finite elements method within COMSOL Multiphysics\textregistered~\cite{Comsol} and imported in the GEANT4 simulation package~\cite{Geant4}, allowing us to study muon motion in the realistic electric field. 
The GEANT4 simulation included the most relevant low energy processes down to eV energies, namely low energy elastic collisions ($\mu^+-$He and muonium$-$He) and charge exchange processes (muonium formation and ionization)~\cite{Bao2014}.
 Cross sections for the elastic collisions and charge exchange were implemented by energy and velocity scaling ~\cite{Senba1989}, respectively, of the proton cross sections from ~\cite{Krstic2006,Nakai1987}.

\begin{figure}[]
	\includegraphics[width=1.0\columnwidth]{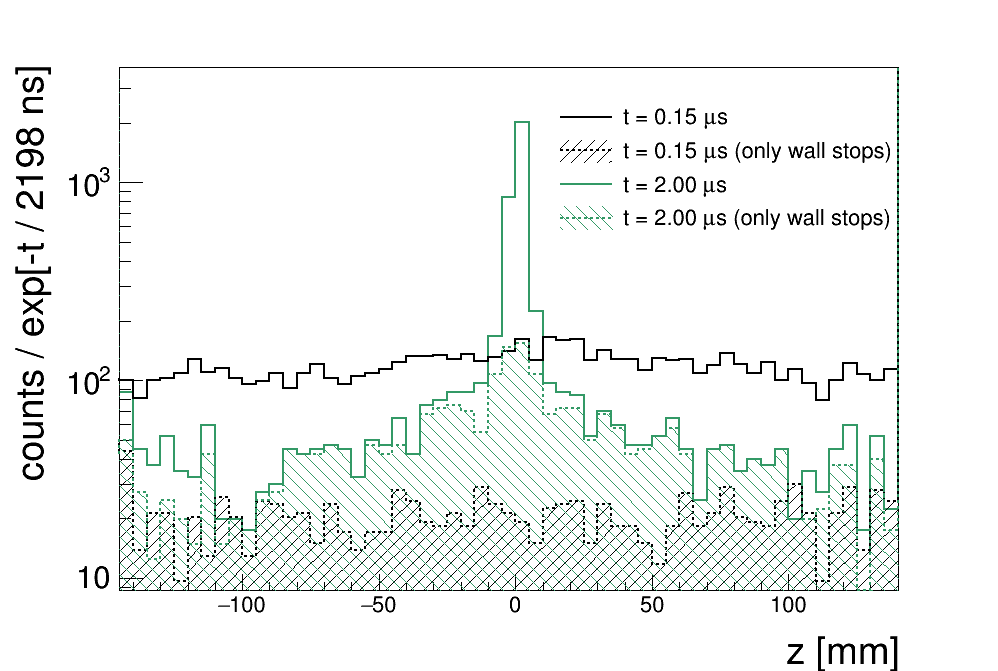}%
		\caption{\label{FigSpaceDistSim}
	Simulated muon distribution along the $z$-axis for various times. The data has been corrected for the finite muon lifetime by multiplying the counts with $e^{t/2198\si{\nano \second}}$.
	As can be seen, for $t=2$~\si{\micro \second} most of the muons are close to the center of the active region.
	The shaded areas show the fraction of the $\mu^{+}$ stopped in the walls of the target. At $t=2$~\si{\micro \second}, the distribution of the $\mu^+$ adhering to the target walls is not flat anymore. The reason for the peak around $z=0$ is that some of the $\mu^+$ are scattered into the wall while drifting towards the center of the target.
	}
\end{figure}

A simulation of the muon distribution inside the target at two different times $t$ is shown in Fig.~\ref{FigSpaceDistSim}. 
The distribution at $t=0.15$~\si{\micro \second} represents the initial muon stop distribution, while the distribution at $t=2$~\si{\micro \second} shows the muon distribution when the longitudinal compression is almost completed. To highlight the muon swarm compression, the number of counts has been scaled with $\exp(t/\tau)$, where $\tau=2198$~\si{\nano \second} is the muon lifetime, in order to compensate for the muon decay.

Under such conditions, at $t=2$~\si{\micro \second}, 63\% of the muons that were in the active region at $t=0.15$~\si{\micro \second} are still within the active region, with 90\% of them already within $z=\pm5$~mm. The 50\% of all the muons in the active region at $t=2$~\si{\micro \second} are already compressed in the center within an even smaller region of $z=\pm1$~mm. Therefore, the number of muons within the region of $z=\pm1$~mm increases by about a factor of 20 between 0.15~\si{\micro \second} and 2~\si{\micro \second} (neglecting muon decay). 

The other 37\% of the muons that were initially in the active region are lost through two main mechanisms: 26\% through muonium formation and 11\% through scattering out of the active region due to low-energy elastic collisions.
The muon bound in the neutral muonium atom is not contained by the 5 Tesla magnetic field, causing it to fly into the walls of the target. The losses due to scattering of $\mu^+$ are only relevant when the muon beam is injected into the longitudinal target at keV energies because the  $\mu^+$ mean free path at these energies can be up to few cm. These losses will therefore be absent in the final setup, where muons enter the longitudinal compression stage from the transverse compression stage at eV energies. Indeed, at such energies, the $\mu^+$ mean free path is sub-mm. Similarly, the losses due to the muonium formation will be strongly reduced in the final setup.

\begin{figure}[]
	\includegraphics[width=1.0\columnwidth]{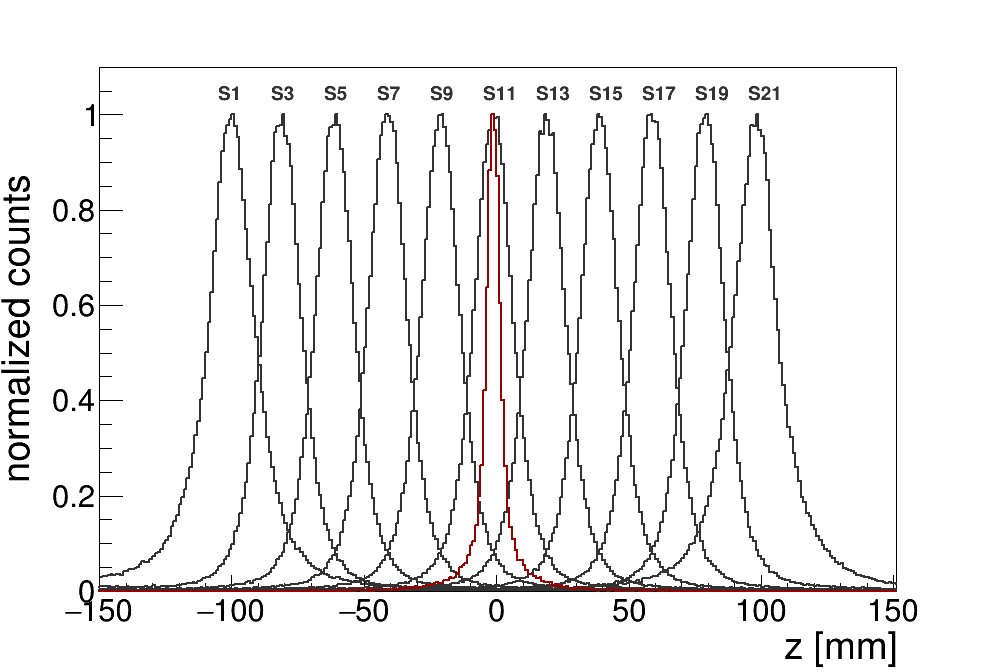}%
	\caption{\label{FigGeomAcc}
	Simulated geometrical acceptance of the scintillators S1-S21 (only every second detector is shown) versus the $z$-position. Shown is the normalized probability that a positron from a $\mu^+$ decay will be detected in the corresponding scintillator. For comparison, the geometrical acceptance of the T1 \& T2 coincidence is also plotted (red line).
	The maximum of the efficiency for each detector was normalized to 1 to highlight the different geometrical resolutions.
	}

\end{figure}

The muon swarm movement was also measured experimentally by placing 21 identical scintillators (S1 to S21) along the target $z$-axis, as shown in Fig.~\ref{FigSetup2014}. The scintillators detected positrons from muon decays and were read out by Silicon Photomultipliers (SiPMs). The acceptance of some of these scintillators as a function of muon decay $z$-position is plotted in the Fig.~\ref{FigGeomAcc}. Each detector has an average geometrical acceptance in $z$-direction of $16.5$~mm (FWHM).

\begin{figure}[]
	\includegraphics[width=1.0\columnwidth]{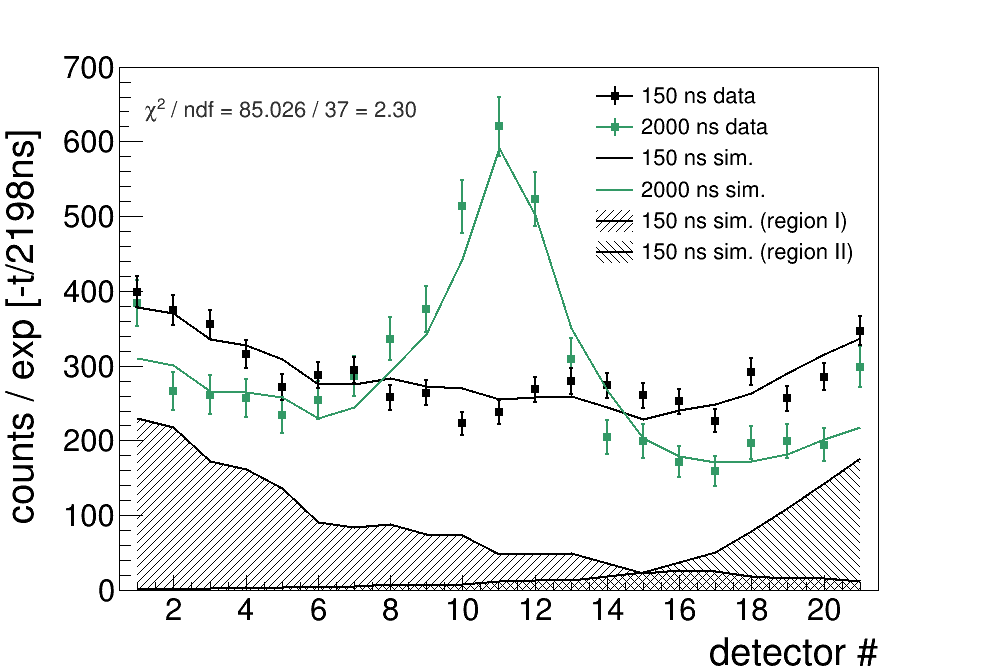}%
	\caption{\label{FigLongSpatialDistribution}Measured (dots) and simulated (lines) positron hits in the 21 scintillators shown in Fig.~\ref{FigSetup2014}, for $t=150~\si{\nano \second}$ and $t=2000~\si{\nano \second}$. The center-to-center distance between two adjacent scintillators is $10$~mm. The simulated data is the same as in Fig.~\ref{FigSpaceDistSim}, but convoluted with the geometrical acceptance of the detectors. The width of the peak at $t=2$~\si{\micro \second} is due to the large geometrical acceptance of the 21 scintillators and does not reflect the width of the muon swarm directly (compare with Fig.~\ref{FigSpaceDistSim}).
	The shaded areas show the simulated contribution to the positron hits from muons stopped in regions I and II.
	The data and the simulations have been corrected for the finite muon lifetime by multiplying the counts with $e^{t/2198\si{\nano \second}}$.}
\end{figure}

The number of measured positrons in each scintillator is presented for $t=150$~\si{\nano \second} (black dots) and $t=2000$~\si{\nano \second} (green dots) in Fig.~\ref{FigLongSpatialDistribution}. Note that the number of counts has been scaled with $e^{t/2198\si{\nano \second}}$ to compensate for the muon decay. 

In Fig.~\ref{FigLongSpatialDistribution} the measured $\mu^+$ distributions (dots) are  compared to the GEANT4 simulations (lines). The simulated number of positron hits in each of the detectors S1-S21 is obtained by convoluting the simulated spatial distribution of the muon swarm at the corresponding time, given in Fig.~\ref{FigSpaceDistSim}, with the detection efficiency of the appropriate detector, given in Fig.~\ref{FigGeomAcc}. These positron hits originate not only from the muons in the active region, but also from muons from the regions where the electric field is not well defined, giving rise to a substantial background.

This background is dominated by positron hits from the $\mu^+$ that stop in the regions I and II (as defined in Fig.~\ref{FigSetup2014}). The background is larger for the detectors placed at the periphery of the active region (S1, S2 and S20, S21), as shown in the Fig.~\ref{FigLongSpatialDistribution} (shaded areas). The shape of the background caused by these muons can be simulated. However, the exact number of $\mu^+$ which stop in the regions I and II depends strongly on the momentum distribution of the initial muon beam, which is not sufficiently well known.

Therefore, two measured distributions along the $z$-axis (for $t=150$~\si{\nano \second} and $t=2000$~\si{\nano \second}) are fitted simultaneously with the sum of 4 contributions:
\begin{enumerate}
	\item Background arising from the region I
	\item Background arising from the region II
	\item Linear background
	\item Simulation of all the $\mu^+$ that stop in the gas (which includes $\mu^+$ in the active region).
  \end{enumerate}
The shape of these 4 contributions is known, under assumption that all the detectors (S1 to S21) have the same detection efficiency. Each of the contributions has to be scaled with a different scaling factor to account for the different stopping probability in region I, region II, gas, and prompt stop at the target lateral walls. 
The additional linear background allows us to account for possible misalignment between the target and the magnetic field axis, which would lead to different numbers of muon wall stops at the position of the various scintillators.

Even though we observe fair agreement between the measurement and the simulation (reduced chi-square $\chi^2_{\rm red}=2.3$, for 37 degrees of freedom), it is difficult to extract precise values of the compression efficiency and of the width of the muon swarm from these measurements, given the limited geometrical resolution of the detectors S1 to S21 and the large background from regions I and II. The relatively large $\chi^2_{\rm red}$ could be attributed to the small variations of the detector efficiencies.

\begin{figure}
	\includegraphics[width=1.0\columnwidth]{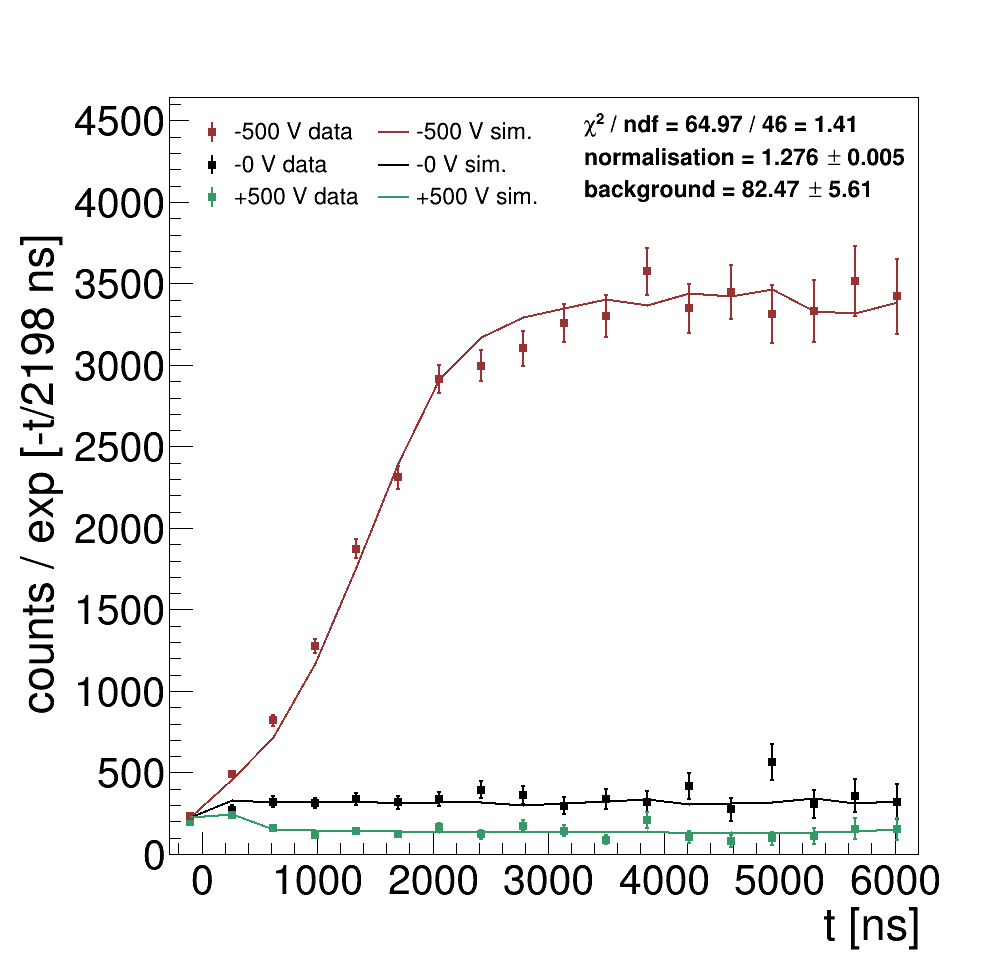}%
	\caption{\label{FigLongTimeSpectraSimulationData}Measured (dots) and simulated (line) time spectra for 5~mbar He gas pressure and potentials of $-500$~V (red), 0~V (black) and $+500$~V (green). 
	The data has been normalized to the number of incoming muons in the entrance detector D1 and fitted simultaneously with only 2 free parameters, a common normalization and a common background.  In total $5\cdot10^8$ muons have been simulated.
	The data and the simulations have been corrected for the finite muon lifetime by multiplying the counts with $e^{t/2198\si{\nano \second}}$.
	}
\end{figure}

In order to better quantify the compression efficiency we turn our attention to the two telescope detectors T1 \& T2 in coincidence that were placed in the center of the target, at $z=0$, as shown in Fig.~\ref{FigSetup2014}. Massive brass shielding all around the target  ensured that coincidence hits in T1 \& T2  originated only from muons decaying within the small region between about $z=\pm3$~\si{\milli \meter} in the center of the target, as shown in Fig.~\ref{FigGeomAcc}.
From the time difference $t=t_1-t_0$ between the positron hit in T1 and T2 in coincidence (at time $t_1$) and the entrance detector, at time $t_0$, a time spectrum can be obtained as shown in Fig.~\ref{FigLongTimeSpectraSimulationData}. The time spectra were recorded for different applied electric potentials.

It can be seen in Fig.~\ref{FigLongTimeSpectraSimulationData} that if no electric field is applied (black points) the number of muons decaying in the center of the target (in T1 \& T2 acceptance region) stays constant (after compensation for the $\mu^+$ decay). 
 When a negative potential (red points) is applied in the center of the target cell, the measured number of counts increases with time. In that case, the stopped muons are attracted towards the potential minimum, so that more muons decay within the acceptance region of T1 \& T2. This means that the muon swarm extent has been decreased in $z$-direction, representing a complementary way to demonstrate longitudinal compression. On the contrary, if the ``wrong" polarity is applied (green points), the muons drift away from the center of the target, out of the acceptance region of T1 \& T2. The very few counts at late times in that case are due to some small background (mostly muons stopping in the wall of the target). The reduction to a nearly background-free measurement represents a major improvement compared with the earlier measurements from 2011~\cite{Bao2014} and compared to the measurement of the $\mu^+$ $z$-distribution of Fig.~\ref{FigLongSpatialDistribution}.

The measurements of Fig.~\ref{FigLongTimeSpectraSimulationData} are compared to GEANT4 simulations. The three simulations ($+$, $-$ and $0$ voltage) are fitted simultaneously to the corresponding measured time spectra. Only two free parameters were used: a common scaling factor and a common flat background. The common scaling factor is needed to remove the uncertainties related to the positron detection and muon stopping efficiencies.
The flat background was included in the fit to account for potential misalignment of the target with respect to the magnetic field axis, which would lead to increased muon stops in the walls. 

The simultaneous fit of the 3 curves has a reduced chi-square $\chi^2_{\rm red} = 1.41$ (for $46$ degrees of freedom). Introduction of additional losses during the compression process, as detailed in the next section, improves the $\chi^2_{\rm red}$ value to $0.95$. Alternatively, a smaller $\chi^2_{\rm red}$ can also be obtained by a minor tuning of the detector acceptance, related to uncertainties in the position, tilt and energy threshold of the T1 \& T2 scintillators.

\section{Additional muon losses?}

As mentioned in the introduction, in the previous 2011 experiment~\cite{Bao2014}, the data quality also suffered from low-energy muon losses due to impurities present in the helium gas, which were caused by the use of Araldite glue and PCB boards in the target construction. This manifested itself by a fast termination of the compression (around $t=0.5$~\si{\micro \second}), because the low-energy muons were captured by the contaminant molecules forming muonic ions or replacing a proton from such a contaminant molecule. 
By implementing a ``chemical capture" rate for muons with energies $\leq10$~\si{\electronvolt} in the simulation, it was possible to roughly mimic this effect. 

In the experiments presented here, care was taken to ensure high gas purity. This was achieved by realizing the target from glass plates, using a low-outgassing glue and purifying the He gas in a cold-trap before feeding it into the target.

\begin{figure}[!h]
	\includegraphics[trim=0cm 0cm 0cm 0cm, clip=true,width=1.0\columnwidth]{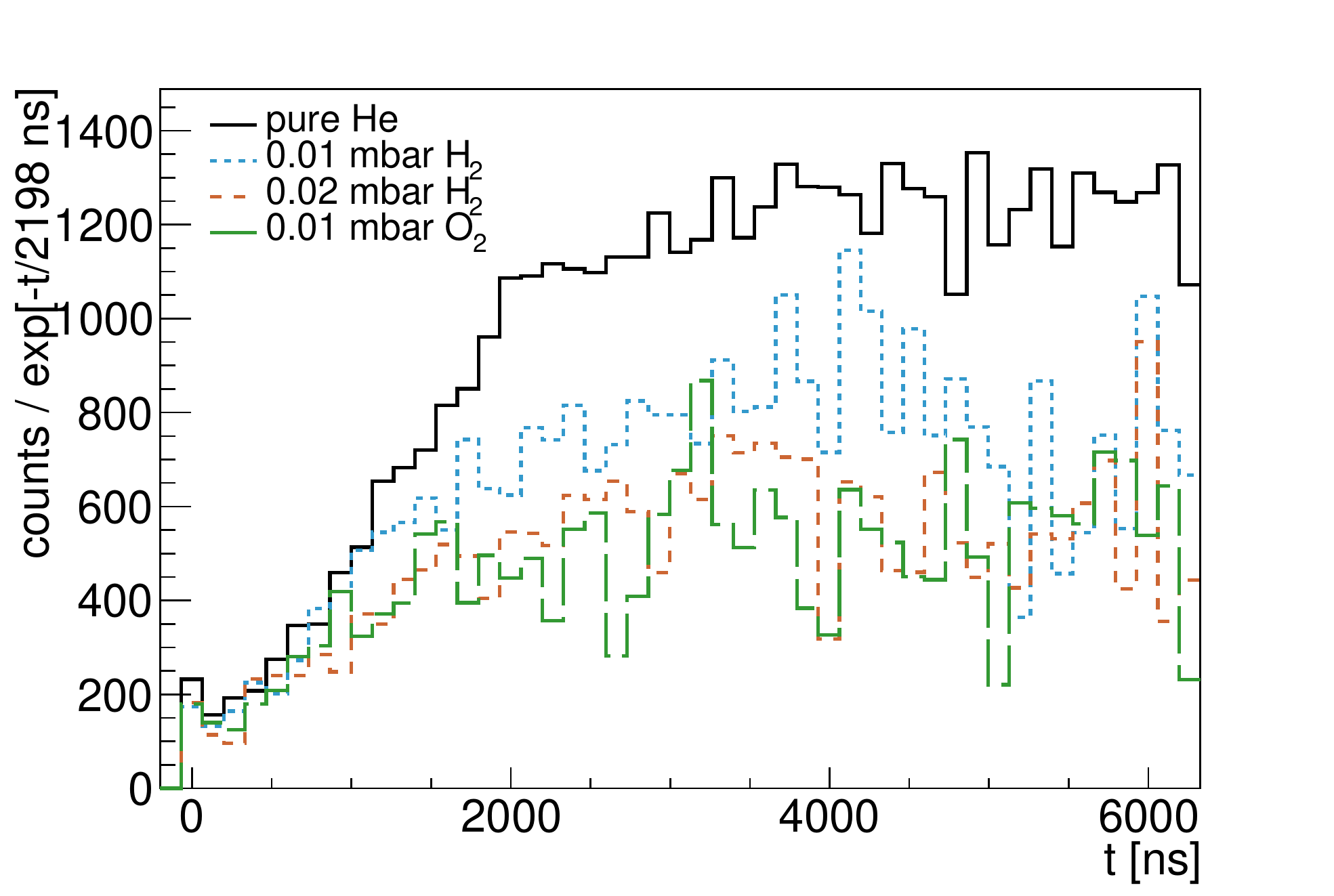}%
	
	\caption{\label{FigImpurities}Measured time spectra for 5~mbar pure He gas, and various admixtures of  contaminants: (black) no additional contaminants, (dashed blue) 0.01~mbar H$_2$, (dashed orange) 0.02~mbar H$_2$, (dashed green) 0.01~mbar of O$_2$. 
	All data were normalized to the number of incoming muons and corrected for the finite muon lifetime by multiplying the counts with $e^{t/2198\si{\nano \second}}$.}
\end{figure}

In order to prove the hypothesis that indeed the compression efficiency is lowered if contaminants are present in the gas, we introduced in a controlled way different amounts of contaminant gases into the pure helium gas. The result of this test is presented in the time spectra of Fig.~\ref{FigImpurities}. It is observed that the compression stops at earlier times when the contaminants are introduced, and consequently the compression efficiency decreases (i.e. the number of muons that are brought to the center of target).

The sensitivity of the measurements of Fig.~\ref{FigLongTimeSpectraSimulationData} to the muon loss mechanisms has been investigated assuming constant (in time) loss rates $R$ during compression, that causes the ``free" muon population to decrease according to $e^{-Rt}$. Various simulations have been performed with $R$ ranging from $R=0$ up to $R=0.5$~\si{\per \micro \second}. The loss rate $R=0.5$~\si{\per \micro \second} would correspond roughly to the measurement with $0.01$~mbar H$_2$ contamination (blue dashed line in the Fig.~\ref{FigImpurities}).

\begin{figure}[]
	\includegraphics[trim=0cm 0cm 0cm 0cm, clip=true,width=1.0\columnwidth]{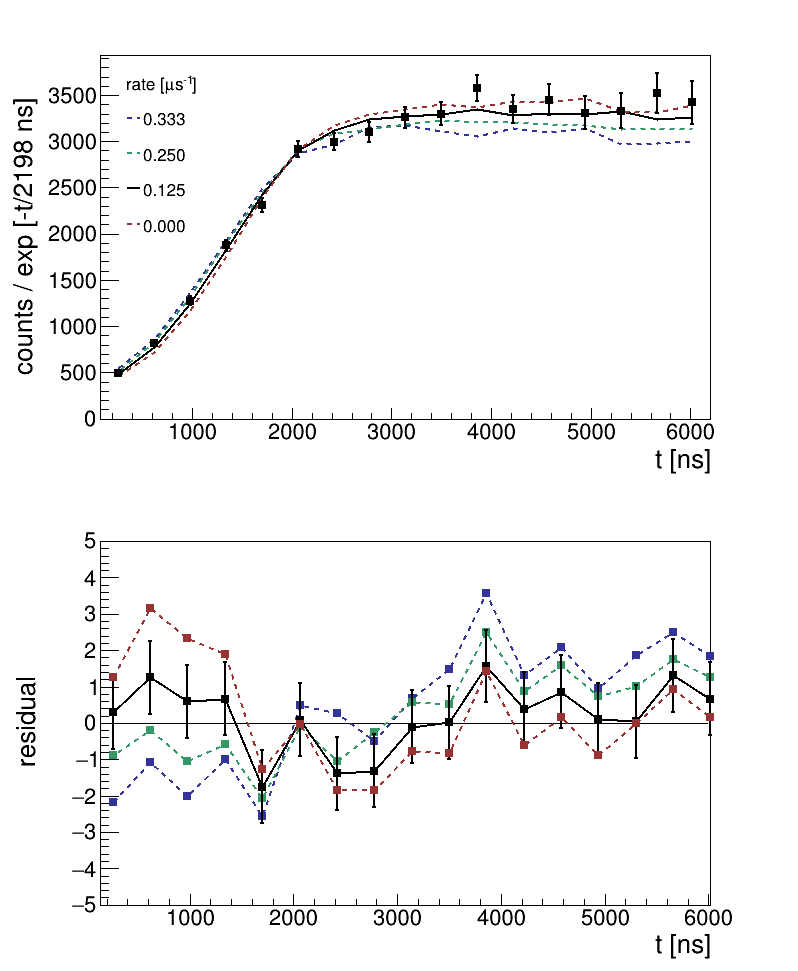}
	\vfill
	\includegraphics[trim=0cm 0cm 0cm 0cm, clip=true,width=1.0\columnwidth]{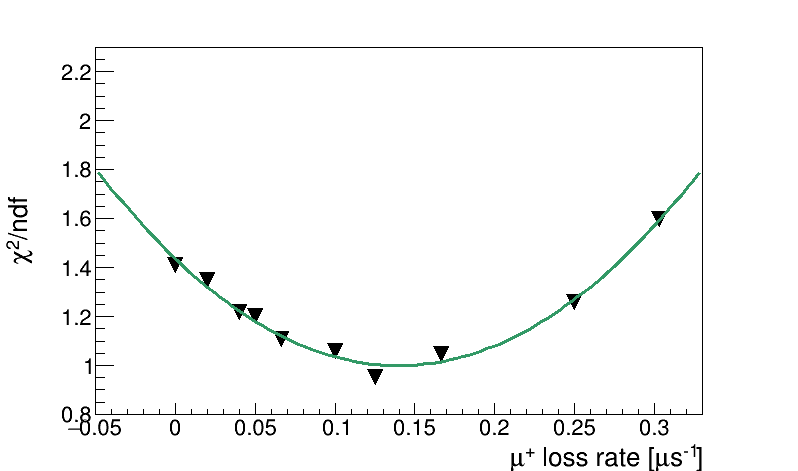}
	\caption{\label{FigVerificationSimulation}(Top) Simulated time spectra for loss rates of $R=(0,0.125,0.25,0.333)$~\si{\per \micro \second} (lines). The $-500$~V measurement is represented by square points. The data and the simulations have been corrected for the finite muon lifetime by multiplying the counts with $e^{t/2198\si{\nano \second}}$.
	(Middle) Residuals (normalized to the data uncertainty) for the time spectra from the top figure. For simplicity, error bars are shown only for the curve with the reduced $\chi^2=0.95$. For other curves error bars are of the similar size.
	(Bottom) Reduced $\chi^2$ as a function of the loss rate $R$. A second order polynomial was fitted to these points (green line).}
\end{figure}

For each loss rate $R$ the simultaneous fit of the simulations for $+$, $-$ and $0$ voltages to the corresponding measurement has been performed (as in Fig.~\ref{FigLongTimeSpectraSimulationData}, but for $R \neq 0$). Figure~\ref{FigVerificationSimulation} (top) shows the fitted time spectra (only for negative voltage) for several loss rates $R$. For each of the loss rates $R$, the $\chi^2_{\rm red}$ between the measurement (only $-500$~V data) and simulations has been calculated and plotted in Fig.~\ref{FigVerificationSimulation} (bottom). A parabola was then fitted to these points (green line). The best agreement (minimum $\chi^2_{\rm red}$) between simulation and measurement is obtained for $R=0.14$~\si{\per \micro \second}, corresponding to an additional loss at $t=2$~\si{\micro \second} of $1-e^{-0.14\si{\per \micro \second} \cdot 2 \si{\micro \second}}=24$\%. The $\chi^2_{\rm red, min}+1$ is obtained for $R=0.35$~\si{\per \micro \second} corresponding to an additional loss at $t=2$~\si{\micro \second} of $51$\%. Therefore, we conclude that the total additional loss after 2~\si{\micro \second} is $= 24^{+27}_{-24}\%$.

If these losses would be caused by capture of the muons by gas impurities, the partial pressure of the impurities would be $1\cdot 10^{-3}-3\cdot 10^{-3}$~mbar. The relation between fitted $R$ and impurity concentration is obtained by fitting the simulation for various loss rates $R$ to the measured time spectra of Fig.~\ref{FigImpurities} with the additional $0.01$~mbar H$_2$ contamination. Because such an impurity level has not been observed in the experiment, we may conclude that the origin of these "effective losses" $R$ cannot be attributed solely to the impurities.

Other explanations, such as uncertainties of the detector acceptance and of the cross sections implemented in the simulation are more favored.

\section{$\vec{E} \times \vec{B}$-Drift}

The next important step towards the realization of the complete muCool device is the demonstration of the so-called $\vec{E}\times\vec{B}$-drift. This drift guides the muons from one compression stage to the next, and is used to extract them finally into vacuum through a small orifice. 

\begin{figure}[]
	\includegraphics[width=0.4\textwidth]{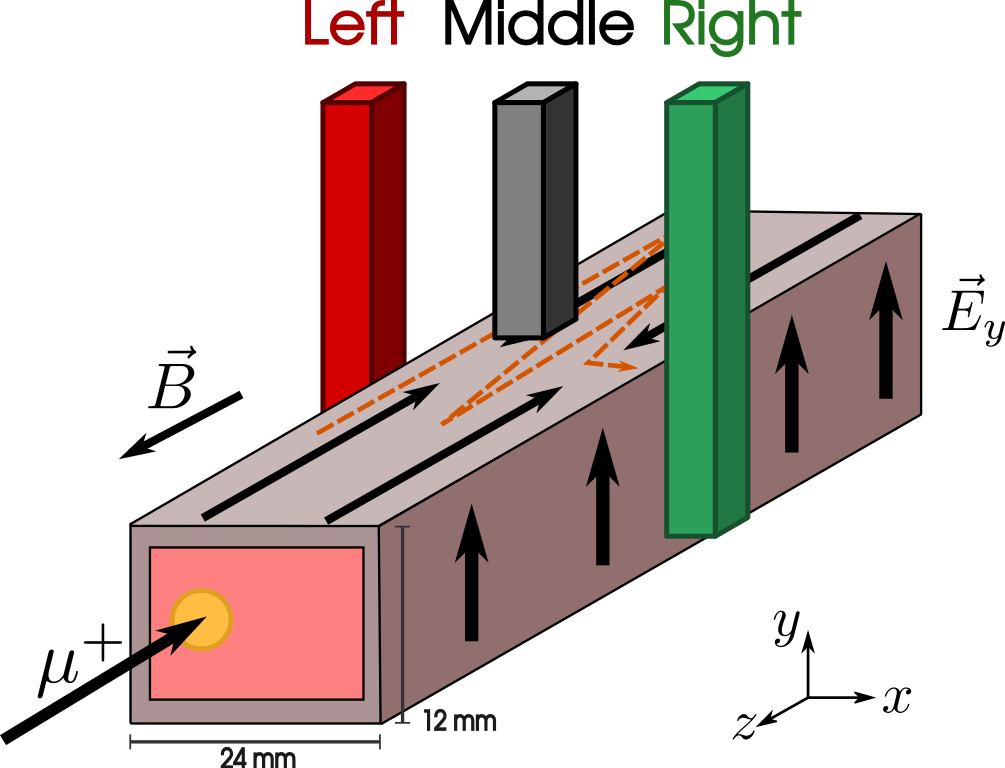}
	\caption{\label{FigDriftSetup} Sketch of the setup to measure the $\vec{E}\times\vec{B}$-drift. The muon beam was injected off-axis to allow the muons to drift a longer distance before hitting the lateral wall. The electric field has a $z$-component for longitudinal compression and a $y$-component to drift the muons in $x$-direction. The three scintillators (``Left", ``Middle", ``Right") are positioned in the center of the target at $z=0$ and monitor the muon swarm movement in $x$-direction.}
\end{figure}

\begin{figure}[]
	\includegraphics[trim=0 0 2cm 2cm,clip=true,width=0.49\textwidth]{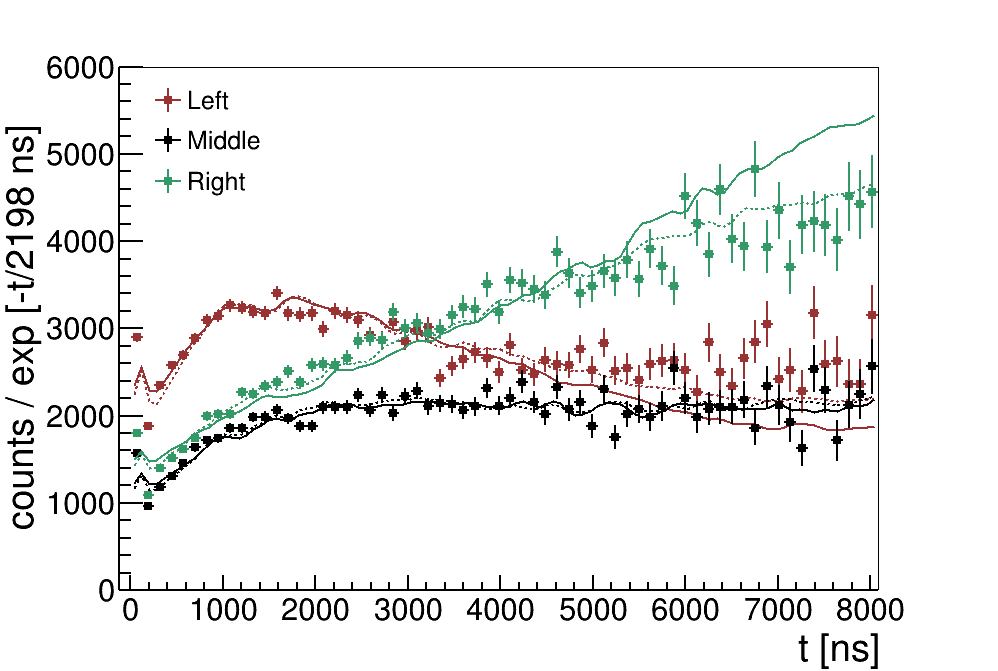}

	\caption{\label{FigDrift}Measured (dots) time spectra for the three scintillators ``Left", ``Middle" and ``Right". The increase and decrease of positron counts in the detectors ``Right" and ``Left", respectively, indicates that the muons are moving towards the right. 
	Simulated time spectra for additional muon loss rate $R=0$ (continuous lines) and $R=0.125$~\si{\per \micro \second} (dashed lines) are fitted to the measured time spectra. The $\chi^2_{\rm red}$ for $R=0$ is $3.44$ for $433$ degrees of freedom. Introducing the additional loss rate $R=0.125$ ~\si{\per \micro \second} improves the $\chi^2_{\rm red}$ to $2.23$.
	Note that the data and the simulations have been corrected for the finite muon lifetime by multiplying the counts with $e^{t/2198\si{\nano \second}}$.
	}
\end{figure}

For this purpose we modified the target cell to generate an electric field with an additional vertical component $E_\text{y}=120$~\si{\volt / \centi \meter}. The non-vanishing central term in Eq.~\ref{EqDriftVelocity} leads to a drift of the muons in $+x$-direction. In total seven scintillators were mounted around the target at $z=0$ along the $x$-direction to monitor this drift. For simplicity, in the sketch of Fig.~\ref{FigDriftSetup} and in the plots of Fig.~\ref{FigDrift} we only show three of them, namely ``Left", ``Middle", ``Right". The target cell was enlarged to a cross section of $24\times12$~\si{\square \mm} and, additionally, the muon beam was injected off-center, as indicated in Fig.~\ref{FigDriftSetup} with the yellow circle. These two modifications allowed $\mu^+$ to drift for longer times before hitting the right wall of the target, thus enhancing the sensitivity of the measured time spectra to the muon drift.

\begin{figure}[]
	\includegraphics[width=0.5\textwidth]{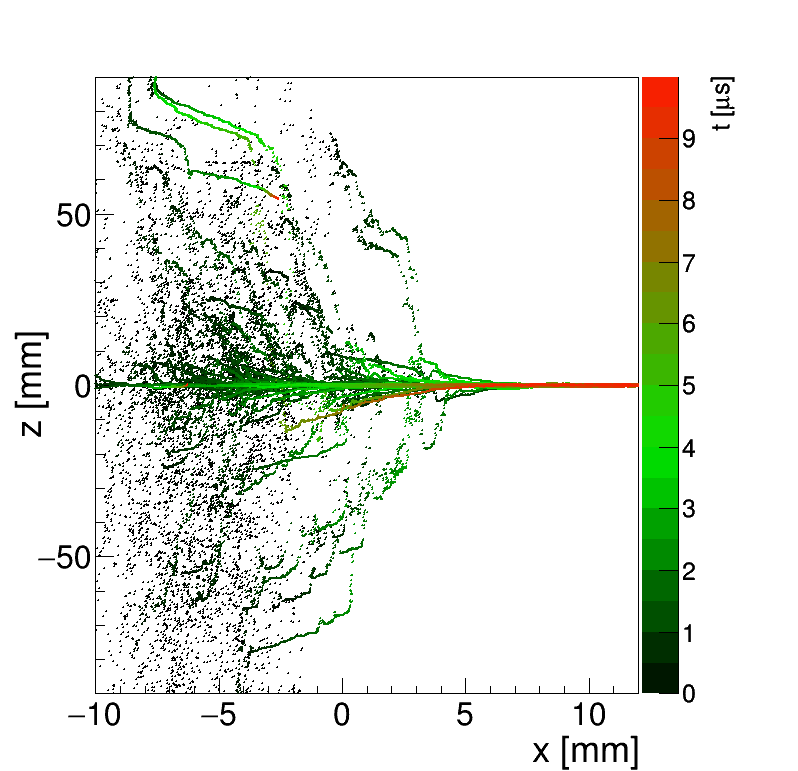}
	
	\caption{\label{FigMuonDrift}Muon positions in the $xz$-plane for various times. The time is given by the color scale. Muon beam centered at $x=-6$~mm with 3~mm radius is stopped uniformly along the $z$-axis. The muons drift in the $+x$-direction while compression occurs in $z$-direction.}
\end{figure}

The simulated spatial distribution of the muon swarm as a function of time is given in Fig.~\ref{FigMuonDrift}. This distribution, convoluted with the corresponding detector acceptance, gives rise to the simulated time spectra shown in Fig.~\ref{FigDrift}, together with the corresponding measurements. 

At early times, the time spectra are dominated by the muon swarm compression in $z$-direction, thus the number of detected positrons increases in all scintillators. However, after about 2~\si{\micro \second}, the number of detected positrons in the scintillator ``Left" decreases, indicating that the muons are moving out of the acceptance region of this scintillator. On the other hand, scintillator ``Right" detects increasingly more positrons. This finding indicates that the muon swarm slowly drifts in $x$-direction towards the prospective point of extraction.

Also in this case, the measured time spectra of all seven ``drift" detectors were fitted simultaneously with the simulation allowing for one common scaling factor, and a different flat background for each detector. A fair agreement between data and simulation has been observed ($\chi^2_{\rm red}=3.44$ for $433$ degrees of freedom).

To study the effect of the additional muon losses on the measured time spectra, the loss rate $R$ has been introduced in the simulation, analogously to the procedure described in the previous section. The obtained time spectra for the various $R$ were then fitted to the data. The best agreement between simulation and the data is obtained for $R=0.125$~\si{\per \micro \second}, consistent with the loss rate $R$ reported in the previous section.
The best fit, which gives a $\chi^2_{\rm red}=2.23$, is shown in Fig.~\ref{FigDrift} (dashed lines).

Even with the additional muon losses introduced, some systematic discrepancies between the data and the simulation still remain. 
Yet the main goal, namely to demonstrate the feasibility of the $\vec{E}\times\vec{B}$-drift, has been achieved. 

The difference between the simulation and the measurement can be attributed either to the simplified modelling of the additional losses (without any energy dependence) or a misalignment of the beam with respect to the magnetic field axis. 

According to the simulation, the drift velocity is about 2~\si{\mm / \micro \second}. This value can be increased in the final setup by increasing the strength of the electric field in $y$-direction.

\section{Conclusions}

The longitudinal compression stage of the muCool device under development at PSI has been demonstrated. An elongated muon swarm of $200$~\si{\milli \meter} length has been compressed to below $2$~\si{\milli \meter} length within 2~\si{\micro \second}. Good agreement between the simulation and the measurement has been observed. Some additional losses which have been parametrized by only one constant $R$ have been introduced to improve the agreement. 

Furthermore, the ability to drift the $\mu^+$ beam in $\vec{E}\times\vec{B}$-direction towards the prospective position of the extraction hole has been demonstrated by performing a measurement with the electric field having also a component perpendicular to the magnetic field. The simulation of the compression and drift towards the extraction hole is in fair agreement with measurements when the small additional loss rate $R$ is introduced in the simulation.

In summary, better agreement between simulations and measurements is achieved by either including small additional losses in the simulation or, more likely, minor tuning of the detector acceptance or minor variation of the cross section of muonium formation, ionization and muonium-He scattering. In any case, even with this additional loss rate $R$, the proposed $\mu^+$ compression efficiency of $10^{-3}$ is attainable.

\begin{acknowledgements}
	The experimental work was performed at the proton accelerator at PSI. We thank the machine and beamline groups for providing excellent conditions.
	We gratefully acknowledge the outstanding support received from the workshops and support groups at ETH Zurich and PSI. Furthermore, we thank F. Kottmann, M. Horisberger, U. Greuter, R. Scheuermann, T. Prokscha, D. Reggiani, K. Deiters, T. Rauber, and F. Barchetti for their help. This work was supported by the SNF grants No. 200020\_159754 and 200020\_172639.
\end{acknowledgements}

\end{document}